\documentclass[fleqn,twoside]{article}
\usepackage{espcrc2mw}
\usepackage{graphicx}
\usepackage{bm}

\newcommand{\beq}{\begin{equation}}
\newcommand{\eeq}{\end{equation}}
\newcommand{\bea}{\begin{eqnarray}}
\newcommand{\eea}{\end{eqnarray}}
\newcommand{\bi}{\bibitem}

\newcommand{\VEC}[1]{{\bf \bm{#1}}} 

\title{ 
Status of Lattice Flavor Physics
} 

\author{
Matthew Wingate\address{Institute for Nuclear Theory,
University of Washington, Seattle, WA 98195-1550}
}
\begin{document}

\begin{abstract}
This talk reviews recent lattice QCD calculations relevant
for quark flavor physics.  Since lattice results must be accurate
and precise to play a definitive role in phenomenology,
the focus is on unquenched results of quantities which can be 
 calculated most reliably.
\end{abstract}

\maketitle

\section{INTRODUCTION}
\label{sec:intro}

In his talk which preceded mine, V.~Lubicz did an excellent
job summarizing the importance of lattice QCD results toward
the goal of understanding quark flavor changing interactions,
both within the CKM framework and beyond the Standard Model
\cite{Lubicz:Lat2004}.  I.~Shipsey gave a similarly inspirational
talk describing the \hbox{CLEO-c} program, which will
measure many quantities which are within reach of precise lattice
predictions~\cite{Shipsey:Lat2004}. 
 That comparison will generate confidence that
lattice results for experimentally inaccessible quantities 
are correct.  Those two talks provide ample motivation for this talk,
in which I attempt to present a snapshot of the current efforts
to contribute to the study of flavor physics.  These are still
the early days of unquenched lattice QCD.
Unquenched simulations are being pushed deeper into the chiral regime,
and systematic uncertainties are receiving increased scrutiny.
Consequently,
this talk should be read as a status report of ongoing research
rather than a summary of completed work.

Accurate, precise results are required.  Before one can claim
to see new physics, the Standard Model result must be solid:
uncertainties need to be controllable.  The lattice
QCD formulation provides an {\it ab initio} method for computing leptonic
and semileptonic decay properties of hadrons.  In practice, 
however, lattice simulations require at least one additional hypothesis.  
Because the light quark masses cannot be set to their physical values,
data obtained with unphysically heavy quarks must be extrapolated
to the physical limit.  Chiral perturbation theory gives the
functional forms for these extrapolations, but this expansion
cannot be applied if the quark masses are too large.  

If one cannot firmly use chiral perturbation theory to extrapolate
simulation results, then one is forced to extrapolate empirically,
estimating the subsequent uncertainty by trying many fit functions.  
This introduces a hypothesis which is one step removed from the 
desired first principles QCD calculation.
Current results based on simulations using Wilson-type fermions 
have used quark masses larger than $m_s/2$, which is on the border
of the chiral regime.  (Domain wall fermions
simulations have been pushed to $m_s/4$.)
It is hard to test these fitting ans\"atze without doing the
lighter quark mass simulations.  Of course the chiral regime
will eventually be reached and, with it, greater theoretical control.

\begin{figure}[t]
\vspace{6.5cm}
\includegraphics{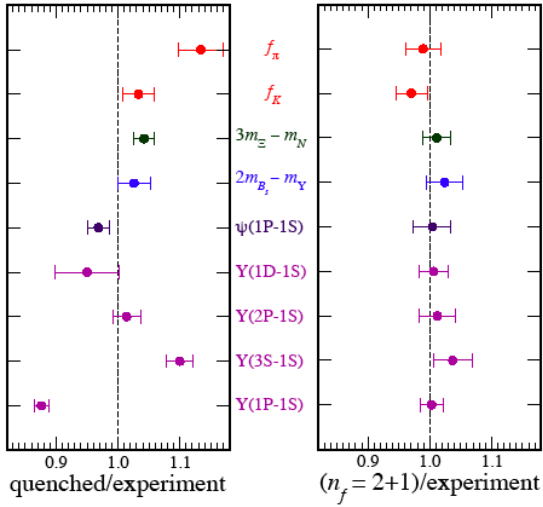}
\caption{\label{fig:ratioplot} Comparison
of quenched and unquenched results to experiment \cite{Davies:2003ik}.}
\end{figure}

Staggered fermions are computationally inexpensive, so lighter
quark masses can be used in simulations. Light sea
quark masses as small as $m_s/10$ have been used to date.  This allows
one to simulate in a region of parameter space where chiral 
perturbation theory converges quickly.  However, there is a price to 
be paid. An algorithmic trick is needed in order to have
3, not 4, active sea quark flavors.  This ``fourth-root'' trick
has not been theoretically justified, so one must hypothesize that
staggered lattice QCD in the continuum limit is QCD.
This hypothesis is testable, and has passed a first round.
Fig.~\ref{fig:ratioplot} shows results for several ``golden,'' or
cleanly-computable, quantities divided by the experimentally
determined value \cite{Davies:2003ik}.  Calculations on the left-hand
side were done in the quenched approximation, and those on the
right-hand side used $2+1$ flavors of dynamical staggered quarks
(the MILC configurations, {\it e.g.}~\cite{Bernard:2001av}).

I believe that these two hypotheses are complementary and both should 
be pursued until neither is necessary.

Due to time and space constraints, some choice had to be made
about which topics to cover and which to omit.  I have given priority
to quantities most cleanly extracted from simulations and
to results which include sea quark effects.
I am unable to cover many interesting and important topics, most
notably $K\to \pi\pi$ matrix elements 
(see {\it e.g.}\ \cite{Ishizuka:2002nm}) and theoretical methods
aimed at reducing systematic uncertainties in future 
calculations (see {\it e.g.}\ \cite{Sommer:2003wm}).

\section{DECAY CONSTANTS}
\label{sec:decayconst}

The decay constant, $f_M$, for a meson $M$ parameterizes the 
leptonic decay matrix element of the weak axial current:
\beq
\langle 0 | A_\mu | M(p) \rangle ~\equiv~ f_M p_\mu \, .
\eeq
Matrix elements of this form are straightforward to compute
in simulations. An additional renormalization calculation is needed
to match the lattice current(s) to the continuum current.
The results presented below rely on perturbative calculations
of the matching coefficients, and in some cases the uncertainty
due to truncating the perturbative expansion is a leading source of
uncertainty.

The $B_s$ and $D_s$ decay constants require no light quark mass
extrapolation, so they are more straightforward to compute than
$f_B$ and $f_D$.  These quantities have been computed using the
$2+1$ flavor MILC configurations.  Using NRQCD for the $b$ quark
we found \cite{Wingate:2003gm}
\beq
f_{B_s} ~=~ 260 \pm 7 \pm 28 ~\mathrm{MeV} \, .
\label{eq:fBsHPQCD}
\eeq
In this work the statistical uncertainty is quoted
first and the total systematic error second, unless noted 
otherwise. 
The systematic error in (\ref{eq:fBsHPQCD}) is 
mostly due to assuming the truncated
$O(\alpha_s^2)$ terms in the perturbative matching contribute with
coefficient $O(1)$.  A calculation of $f_{B_s}$ on the same configurations
using relativistic heavy quarks is underway by FNAL/MILC/HPQCD.

\begin{figure}[t]
\vspace{3.8cm}
\includegraphics{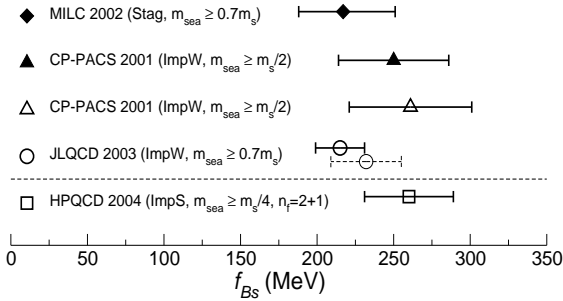}
\caption{\label{fig:fBs_summary} Summary of unquenched calculations 
of $f_{B_s}$ (bold points)
\cite{Bernard:2002pc,AliKhan:2000eg,AliKhan:2001jg,Aoki:2003xb,Wingate:2003gm}.  
Solid points used the Fermilab heavy quark formulation
and open points used NRQCD.  All points above the dashed line have
$n_f=2$.  The sea quark action and masses are noted in parentheses.
The dashed point below the JLQCD result is discussed in the text.}
\end{figure}

Several calculations of $f_{B_s}$ have been done with $n_f=2$.
Although the strange sea quark is quenched, one argues that
strange quark loop effects are small corrections.  This may be
true, but it takes us another step away from the {\it ab initio}
calculation we desire.  Results are summarized in Fig.~\ref{fig:fBs_summary}.
The only notable difference is between the JLQCD result \cite{Aoki:2003xb}
and (\ref{eq:fBsHPQCD}).
Before one attributes this to a difference between 2 and 3 flavor
calculations in general, I think further investigation into the dependence
on light sea quark mass is necessary.  The lesson of Fig.~\ref{fig:ratioplot}
is that
light sea quarks are needed for $f_K$, $f_\pi$, and $\Upsilon$ splittings,
for example, to agree with experiment.  JLQCD find that if the lattice
spacing is set using $f_K$, then $m_\rho$ and $r_0$ agree with 
experiment and phenomenology, respectively.  However, it is also important
that the $\Upsilon$ spectrum be reproduced.  Comparing lattice spacings
set with different quantities, CP-PACS \cite{AliKhan:2001jg} 
found $a_\rho/a_\Upsilon \approx 1.15$ on a coarse lattice and
$a_\rho/a_\Upsilon \approx 1.20$ on a fine lattice.
The JLQCD lattices are similar in quark mass and lattice spacing, so
I speculate that one would find similar results for 
$a_\rho/a_\Upsilon$ (and thus for $a_{f_K}/a_\Upsilon$)
on the JLQCD lattices.  If I include an uncertainty due to lattice
spacing ambiguity in the same manner as \cite{AliKhan:2001jg}, then
I obtain the dashed point in Fig.~\ref{fig:fBs_summary}.
Therefore, I conclude that difference between 2 and 2+1 flavor
results for $f_{B_s}$ cannot be resolved yet.  On the other hand,
there is no advantage to averaging them together, so I quote the 2+1
flavor result as the best current estimate.

\begin{figure}[t]
\vspace{4.5cm}
\includegraphics{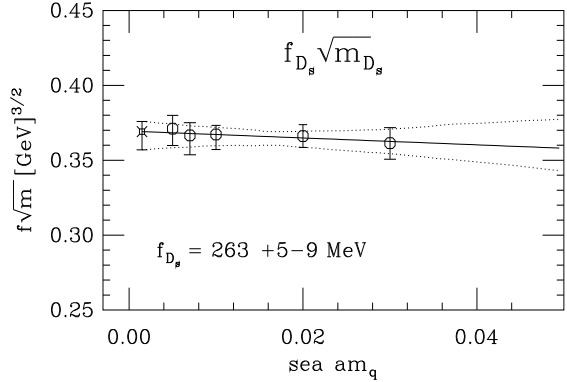}
\caption{\label{fig:fDs-vs-msea} Unquenched calculation  ($n_f=2+1$) 
of $f_{D_s}\sqrt{m_{D_s}}$ vs.\ sea quark mass \cite{Simone:Lat2004}.
(The bare strange quark mass is 0.04 in lattice units.)}
\end{figure}

Lattice NRQCD breaks down as the heavy quark mass is lowered 
to near the inverse lattice spacing.  Consequently extrapolations
must be performed to the charm sector, incurring a large systematic
uncertainty.  It is preferable to simulate directly at the charm
quark mass using relativistic fermions.  This can be accomplished
on very fine lattices, treating the charm quark just like the strange
and light quarks, or on lattices with any reasonable spacing using
the Fermilab formulation~\cite{El-Khadra:1996mp}.
A preliminary calculation of $f_{D_s}$ on the MILC lattices using
the latter method was presented here~\cite{Simone:Lat2004}.  
Fig.~\ref{fig:fDs-vs-msea} shows $f_{D_s}\sqrt{m_{D_s}}$ as a 
function of sea quark mass.  They find
\beq
f_{D_s} ~=~ 263 ~\begin{array}{l}+5\\-9\end{array} 
\pm 24 ~\mathrm{MeV} \, .
\label{eq:fDsFNAL}
\eeq
The quoted systematic uncertainty is mostly due to higher order
in the heavy quark expansion.  This calculation computes part of the
matching nonperturbatively \cite{El-Khadra:1997hq}; 
the remaining 1-loop correction is
small, so the quoted $O(\alpha_s^2)$ uncertainty is negligible at this
level.

\begin{figure}[t]
\vspace{4.5cm}
\includegraphics{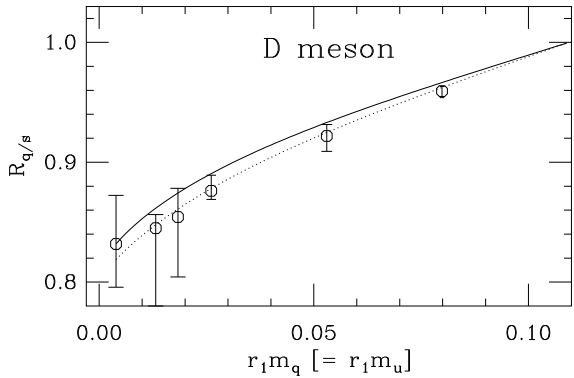}
\caption{\label{fig:fD-fit-full-QCD} Unquenched calculation  ($n_f=2+1$) 
of $f_{D}\sqrt{m_D}/(f_{D_s}\sqrt{m_{D_s}})$~\cite{Simone:Lat2004}. 
 The global fit (dotted curve) to the data points, and to
partially quenched data (not shown), includes taste-breaking $O(a^2)$ effects.
The solid curve is obtained by subtracting these effects and yields
the result in the text.}
\end{figure}

The $D$ decay constant has also been computed in \cite{Simone:Lat2004}.  
Computations were done at many points in 
the partially quenched parameter space, allowing a global fit
to PQ$\chi$PT including finite lattice spacing effects~\cite{Aubin:2004xd}.
Fig.~\ref{fig:fD-fit-full-QCD} shows $f_D\sqrt{m_D}/(f_{D_s}\sqrt{m_{D_s}})$ 
along the fully unquenched line of partially quenched parameter space;
other plots with $m_{\rm sea}$ held fixed were shown too 
\cite{Simone:Lat2004}.  The extrapolated result obtained is
\beq
\frac{f_{D_s}\sqrt{m_{D_s}}}{f_D\sqrt{m_D}} ~=~ 1.20 \pm 0.06 \pm 0.06 \, .
\label{eq:xiD}
\eeq
The error due to truncating the heavy quark expansion cancels in 
the ratio, so the leading systematic uncertainty is from the
chiral fits.  Combining (\ref{eq:fDsFNAL}) and (\ref{eq:xiD}), they find
\beq
f_D ~=~ 224\begin{array}{l} +10\\-14\end{array} \pm 21 ~\mathrm{MeV}
\eeq
where the dominant systematic uncertainty is from the heavy quark
expansion.

\begin{figure}[t]
\includegraphics[width=6.5cm,angle=-90]{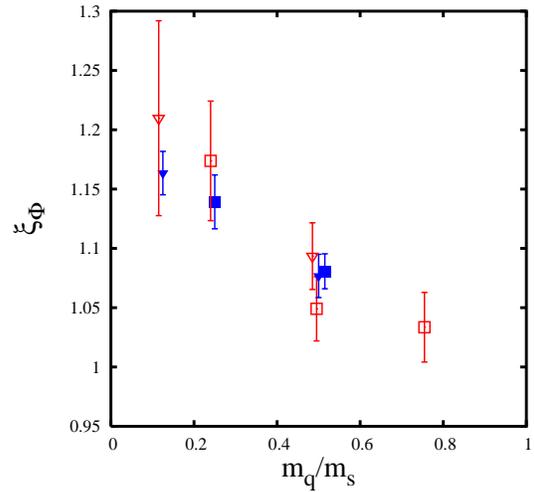}
\caption{\label{fig:xiBGray} $\xi$ vs. valence light quark mass 
in units of physical strange quark mass \cite{Gray:2004hd}.  
The open symbols used point sources and sinks while the solid 
symbols utilized smearing for the heavy quark propagator. Different 
symbols correspond to different sea quark masses.}
\end{figure}

Work by the FNAL/MILC collaboration on the $B$ decay constant using
relativistic fermions is underway.  Since presenting some NRQCD results
last year \cite{Wingate:2003ni}, our efforts have been devoted to
improving the signal-to-noise ratio for $f_B$.  Impressive improvement
was seen by using smeared sources and sinks for the heavy quark
\cite{Gray:2004hd}.
Fig.~\ref{fig:xiBGray} shows the ratio $\xi\equiv f_{B_s}\sqrt{m_{B_s}}/
(f_B\sqrt{m_B})$ plotted vs.\ valence light quark mass.
Current efforts will now be directed toward obtaining data at enough masses 
to permit a PQ$\chi$PT fit as was presented for $f_D$.  There are existing
unquenched calculations which quote results for $f_B$ based on 
extrapolations using $m_{\rm sea} \ge m_s/2$ ({\it e.g.}~\cite{Aoki:2003xb}).
However, due to the importance of this quantity for CKM constraints
and to the possibility of previously unforeseen large curvature at 
small masses \cite{Yamada:2001xp,Kronfeld:2002ab}, I prefer to wait
until the chiral extrapolation is under firmer control before
quoting a result and uncertainty.

The simulations for heavy-light decay constants using the MILC lattices
have been done using 1 lattice spacing and 1 volume so far.
Finite lattice spacing errors have been estimated by assuming
terms neglected in the low energy Symanzik effective theory
appear with $O(1)$ coefficients.  Simulations with finer lattices
are underway to check this estimate.  Finite volume effects were
recently studied in heavy meson chiral perturbation
theory \cite{Arndt:2004bg} and 
estimated to be small compared to the other current uncertainties
quoted above.  However, they could contribute when chiral
extrapolations of $\xi$ are further along.

The predictions from heavy meson chiral perturbation theory are 
crucial ingredients for extrapolating lattice data to the physical
light quark masses.  At small quark masses, chiral logarithms 
become important.  For example, in continuum heavy meson $\chi$PT
$\xi$ has a term which goes like 
$(1+3g^2) m_\pi^2 \ln(m_\pi^2/\Lambda^2)$ times a known constant, where
$g$ is the $B-B^*-\pi$ coupling.  Lattice data for this ratio
will not be able to precisely computed $g$ because simulations are
done at masses where the logarithm is small.  However, a large
uncertainty for $g$ will eventually contribute a significant 
error to $\xi$.  Therefore, it is important to determine $g$
directly.  First results, in the quenched approximation, were published
in \cite{Abada:2002xe} (for the charm sector) and \cite{Abada:2003un}
(for the bottom sector).  It is important to repeat these calculations
with 3 dynamical flavors, since only then are the low energy constants
of chiral perturbation theory physical.


The $f_\pi$ and $f_K$ decay constants have been computed
very precisely on the MILC configurations 
\cite{Aubin:2004fs,Aubin:2004he}.  As pointed out in \cite{Marciano:2004uf},
their result for the ratio
\beq
f_K/f_\pi ~=~ 1.210 \pm 0.004 \pm 0.013
\eeq
can be combined with experimental measurements of the decay rates
for $\pi$ and $K$ to $\mu\overline{\nu}_\mu$ to yield
\beq
|V_{us}| ~=~ 0.2219 \pm 0.0026 
\label{eq:VusMILC}
\eeq
where the error is dominated by lattice uncertainties.
This result is in agreement with, and has the same uncertainty as,
the 2004 PDG average
$|V_{us}| = 0.2200\pm 0.0026$ obtained from semileptonic $K$ decay
\cite{Eidelman:2004wy}.

\section{NEUTRAL $\bm{B}$ MIXING}
\label{sec:Bmixing}

Measurements of the mass and width differences in 
neutral $B$ and $B_s$ mixing can give us insight into flavor changing 
physics.  Accurate computations of hadronic matrix elements
of the relevant 4-quark operators are necessary ingredients in
order to extract the high energy parameters.  In the Standard Model,
one needs to compute
\beq
\langle \overline{B^0} | (\overline{b}d)_{V-A}(\overline{b}d)_{V-A}
|B^0 \rangle \equiv \frac{8}{3} f_{B}^2 m_{B}^2 \mathrm{B}_{B}
\eeq
to obtain $|V_{td}|$ from $\Delta m_d$.  Note that the matrix
element is parameterized by the factor $\mathrm{B}_B$ in a normalization
such that $B_B=1$ in the vacuum saturation approximation.  The 
corresponding calculation for the strange $B^0_s$, combined with
the experimental lower bound for $\Delta m_s$, yields and even
tighter constraint on $|V_{td}|$.

JLQCD results \cite{Aoki:2003xb} are still the state-of-the-art.  
They quote, in the $\overline{\rm MS}$ scheme at $\mu = m_b$
\bea
\mathrm{B}_B~\!\! &\!=\!& 0.836 (27)_{\rm stat} 
{ +0 \choose -27}_{\chi} \!(56)_{\rm EFT} \\
\mathrm{B}_{B_s}\!\! &\!=\!& 0.850 (22)_{\rm stat} { +18\choose -0}_{\chi} 
\!(57)_{\rm EFT}{+5\choose -0}_{m_s} \, .
\eea
One expects the B factors to be less sensitive to chiral logarithms
since their coefficient is much smaller than for the decay constant.
Nevertheless a calculation using light staggered sea quark masses
with $n_f=2+1$ is underway \cite{Gray:2004hd}.

The $B_s$ width difference is dominated by loop diagrams with
a charm quark, so a measurement of $\Delta \Gamma_{B_s}$ would
determine $|V_{cb}^* V_{cs}|^2$.  For this calculation another
matrix element is needed in addition to $\mathrm{B}_{B_s}$, the scalar bag
factor:
\beq
\langle \overline{B^0_s} | (\overline{b}s)_{S-P}(\overline{b}s)_{S-P}
|B^0_s\rangle \equiv
-\frac{ 5 f_{B_s}^2 m_{B_s}^4 \mathrm{B}_{S_s} }{3(m_b+m_s)^2} \, .
\eeq
The best result to date is $\mathrm{B}_{S_s}(m_b) = 0.86(3)(7)$
from \cite{Yamada:2001xp} with $n_f=2$ improved Wilson quarks.
This computation is being done on the MILC lattices, but the
matching calculation needs to be completed.  

Even for physics beyond the Standard Model, lattice calculations
of hadronic matrix elements are needed to connect the experimental
measurements to theoretical parameters.  For example, the complete
set of five 4-quark operators for $B^0-\overline{B^0}$ mixing
have been computed in the quenched approximation~\cite{Becirevic:2001xt}.

\section{SEMILEPTONIC DECAYS}
\label{sec:SLdecay}

Below we discuss how lattice calculations of the form factors
governing semileptonic pseudoscalar states can be used to compute
CKM matrix elements.  The
form factors parameterize the relevant matrix element of the weak current;
{\it e.g.}\ for $B\to \pi\ell\nu$,
\bea
\langle \pi | V_\mu | B \rangle \equiv &\!\!\!\!\!\! f_0(q^2) 
& \!\!\!\!\!\! \frac{m_B^2 - m_\pi^2}
{q^2} q_\mu \nonumber ~+~ \\
 & \!\!\!\!\!\! f_+(q^2) & \!\!\!\!\!\!\left(p_\pi
+ p_B - \frac{m_B^2 - m_\pi^2}{q^2} q \right)_\mu  
\eea
where $q$ is the momentum transferred to the lepton pair.
This matrix element can be computed on the lattice for a range of
$q^2$; one cannot presently go to small $q^2$ without introducing
lattice artifacts as $|\VEC{p}_\pi| \to 1/a$.  
\beq
\frac{1}{|V_{ub}|^2}\frac{d\Gamma}{dq^2} ~=~ \frac{G_F^2}{24\pi^3}
p_\pi^3 \left|f_+(q^2)\right|^2
\eeq
can be integrated over $q^2$ using lattice data for the right-hand side.
Then the partial width, determined by experiment, can be inserted, yielding
$|V_{ub}|$.  

\begin{figure}[t]
\vspace{6.2cm}
\includegraphics{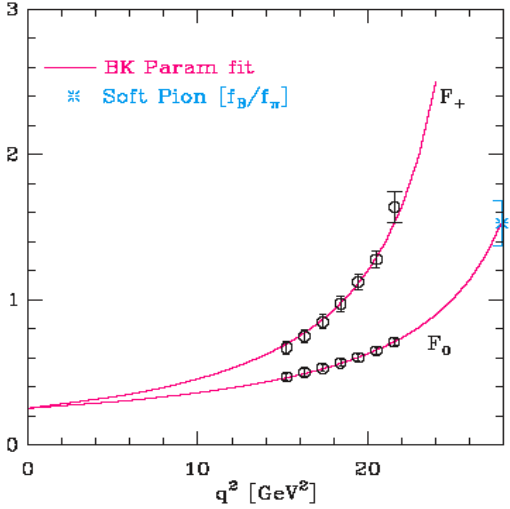}
\caption{\label{fig:B2piFF} $B\to\pi\ell\nu$ form factors on the MILC
configurations with NRQCD \cite{Shigemitsu:2004ft}.}
\end{figure}

A preliminary result using NRQCD on MILC configurations (with
$m_{\rm sea}=m_s/4$) was presented
here \cite{Shigemitsu:2004ft}.  Fig.~\ref{fig:B2piFF} shows $f_+$ and $f_0$
vs.\ $q^2$ after chiral extrapolation of the valence quark mass.  
The data are well-fit by an ansatz which satisfies the kinematic
constraint $f_+(0)=f_0(0)$; the heavy quark scaling laws
$f_+(q^2_{\rm max}) \propto \sqrt{m_B}$, $f_0(q^2_{\rm max})
\propto 1/\sqrt{m_B}$ and $f_{+,0}(0) \propto m_B^{3/2}$;
and includes the $B^*$ pole and an effective second pole
\cite{Becirevic:1999kt}.
Integrating over the entire range of $q^2$ and taking branching
ratios from CLEO \cite{Athar:2003yg} we obtain
\beq
|V_{ub}| = (3.86 \pm 0.32_{\rm expt} \pm 0.58_{\rm latt}) \times 10^{-3} \, .
\label{eq:VubNRQCDall}
\eeq
CLEO has binned their data, so we can perform the
integration over the restricted range 
$16~\mathrm{GeV}^2 \le q^2 \le q_{\rm max}^2$ to find
\beq
|V_{ub}| = (3.52 \pm 0.73_{\rm expt} \pm 0.44_{\rm latt} )\times 10^{-3}  \, .
\label{eq:VubNRQCDpartial}
\eeq
Note that this last step decreases the lattice uncertainty but more
than doubles the experimental error.  The moving NRQCD formulation
\cite{Foley:2002qv,Foley:Lat2004}
promises to permit simulations in an inertial frame moving with
respect to the $B$, so that one can satisfy $|\VEC{p}_\pi| \ll 1/a$ 
even with small $q^2$.

Another preliminary result,
\beq
|V_{ub}| = (3.0 \pm 0.6_{\rm expt} \pm 0.4_{\rm latt})\times 10^{-3} \,,
\label{eq:VubFNAL}
\eeq
was reported here \cite{Okamoto:2004xg}.  This is a fully unquenched
calculation on the MILC lattices with the Fermilab heavy quark formulation,
where the integration uses the same restricted range of $q^2$.
One finds nice agreement between this result and the partially
quenched NRQCD results above.  As the authors of \cite{Shigemitsu:2004ft}
and \cite{Okamoto:2004xg} extend their results on the set of MILC
configurations, results will be highly correlated and a joint
analysis would be the best way to extract a single result.  At present
the results (\ref{eq:VubNRQCDpartial}) and (\ref{eq:VubFNAL}) share
configurations at only one value of $(m_{\rm val},m_{\rm sea})$.
For the purpose of quoting a single number here, I average the
central values in  (\ref{eq:VubNRQCDpartial}) and (\ref{eq:VubFNAL}), but
just carry over the same experimental and lattice errors
\beq
|V_{ub}| ~=~ (3.3 \pm 0.7_{\rm expt} \pm 0.4_{\rm latt} ) \times 10^{-3} \,.
\eeq
The larger experimental error does not let us off the hook; extending
the lattice results more accurately to lower $q^2$ will reduce both
uncertainties, as (\ref{eq:VubNRQCDall}) indicates.

\begin{figure}[t]
\vspace{6.7cm}
\includegraphics{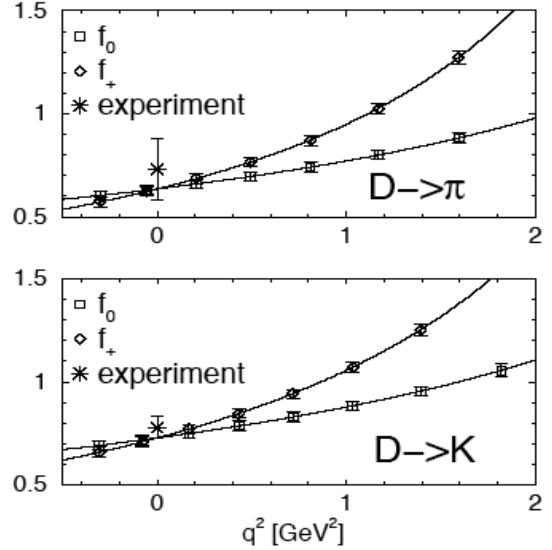}
\caption{\label{fig:DFF} $D\to\pi\ell\nu$ and $D\to K\ell\nu$ 
form factors on the MILC configurations with the Fermilab formulation
\cite{Aubin:2004ej}.}
\end{figure}

\begin{figure}[t]
\vspace{3.7cm}
\includegraphics{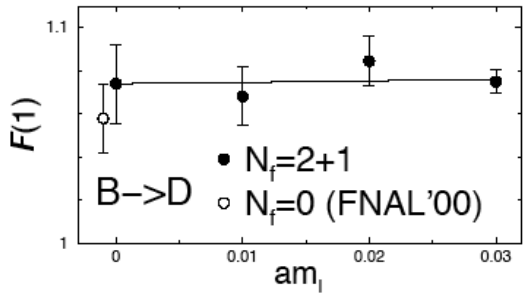}
\caption{\label{fig:BtoDFF} $B\to D\ell\nu$ form factor at
zero recoil vs.\ sea quark mass (in lattice units; $am_s=0.04$) 
\cite{Okamoto:2004xg}.}
\end{figure}

Finalized results for the $D\to\pi\ell\nu$ and $D\to K\ell\nu$
form factors
were presented here \cite{Okamoto:2004xg,Aubin:2004ej}.  These used
the Fermilab heavy quark formulation and fully unquenched light and
strange quarks.
The form factors and their $q^2$ dependence are shown in 
Fig.~\ref{fig:DFF}.  Note the entire range of $q^2$ can be studied
in the $D$ rest frame without the pion having large spatial momentum
in lattice units.  Integrating the fits yields
\bea
|V_{cd}| &=& 0.239 ~(20)_{\rm expt} 
~(10)_{\rm stat} ~(24)_{\rm sys} 
\label{eq:Vcd} \\
|V_{cs}| &=& 0.969 ~ (24)_{\rm expt}
~(39)_{\rm stat} ~(94)_{\rm sys} \, .
\label{eq:Vcs}
\eea
The leading lattice systematic uncertainty is due to
the neglect of higher order terms in the heavy quark expansion.

Ref.\ \cite{Okamoto:2004xg} also presented a preliminary
unquenched result for the
$B\to D\ell\nu$ form factor evaluated at zero recoil,
${\cal F}_{B\to D}(w=1) = 1.074(18)(16)$.  Fig.~\ref{fig:BtoDFF}
shows the light quark mass dependence of the form factor.
Combining this result with experiment gives
\beq
|V_{cb}| ~=~ 0.038 \pm 0.006_{\rm expt} \pm 0.001_{\rm latt}\, .
\label{eq:Vcb}
\eeq
The 3 results (\ref{eq:Vcd}), (\ref{eq:Vcs}), and (\ref{eq:Vcb})
satisfy CKM unitarity:
\beq
\left( |V_{cd}|^2 + |V_{cs}|^2 + |V_{cb}|^2\right)^{1/2} = 1.00(2)(10)
\eeq
where the first error is experimental and the second is from lattice.
This is the first time such a test could be done using only 
inputs from experiment and unquenched lattice QCD.

\begin{figure}[t]
\vspace{6.7cm}
\includegraphics{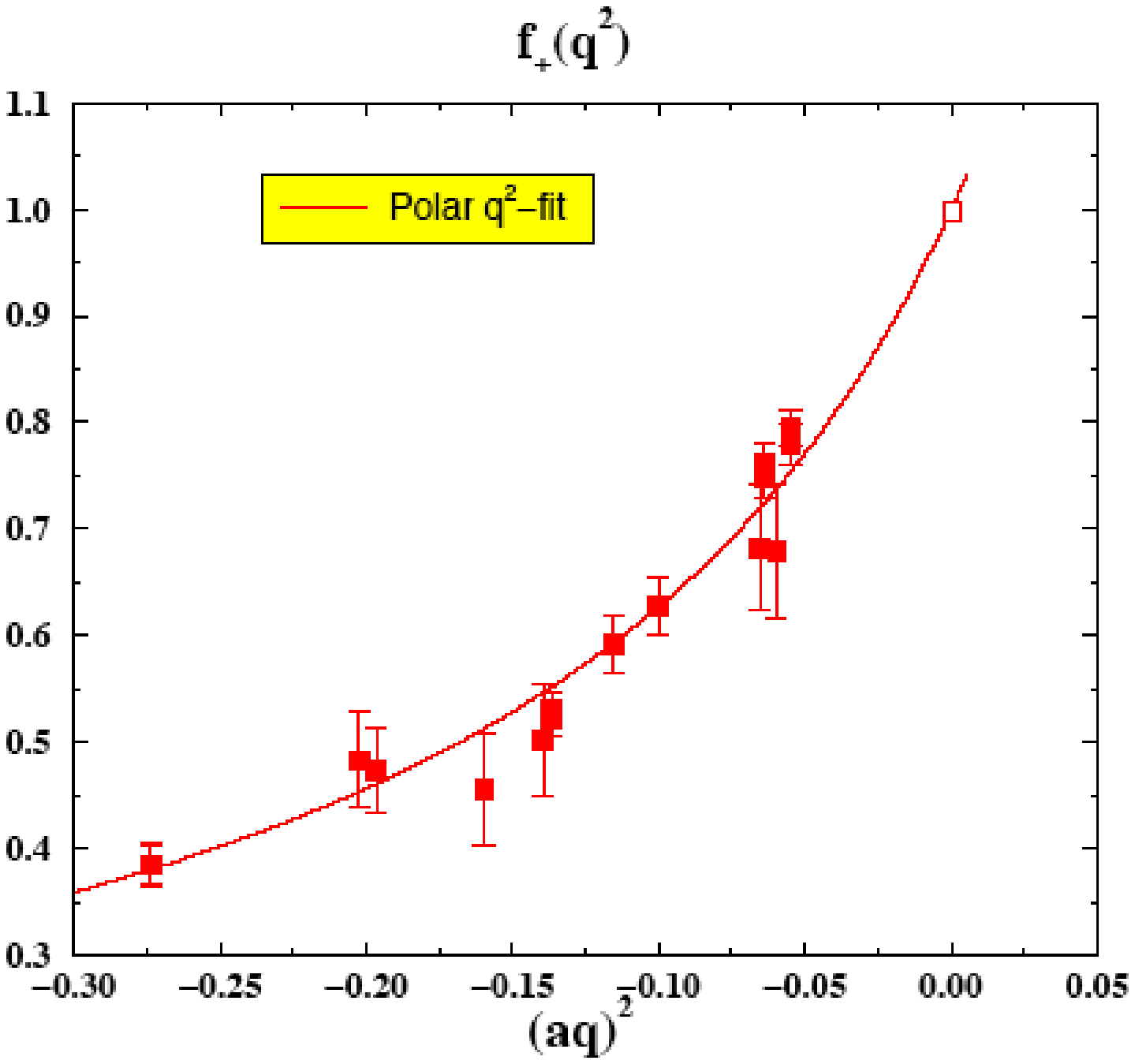}
\caption{\label{fig:K2pi} Quenched $K\to\pi\ell\nu$ form factor
\cite{Becirevic:2004ya}.}
\end{figure}

Progress was made recently in calculating the $K\to\pi\ell\nu$ form 
factors \cite{Becirevic:2004ya}.  Utilizing and extending the double ratio 
method developed for $B\to D\ell\nu$ decays \cite{Hashimoto:1999yp},
the form factors can be computed more precisely than before.  A
plot of $f_+(q^2)$ in the quenched approximation is reproduced in
Fig.~\ref{fig:K2pi}.  Using the value extrapolated to $q^2=0$ with 
experimental input yields 
\beq
|V_{us}|^{\rm quenched} ~=~ 0.2202 \pm 0.0025 \, .
\eeq
This precision is comparable to other methods, in particular
(\ref{eq:VusMILC}), so it would be profitable to use this method
in an unquenched calculation.

\section{NEUTRAL $\bm{K}$ MIXING}
\label{sec:Kmixing}

\begin{figure}[t]
\vspace{6.7cm}
\includegraphics{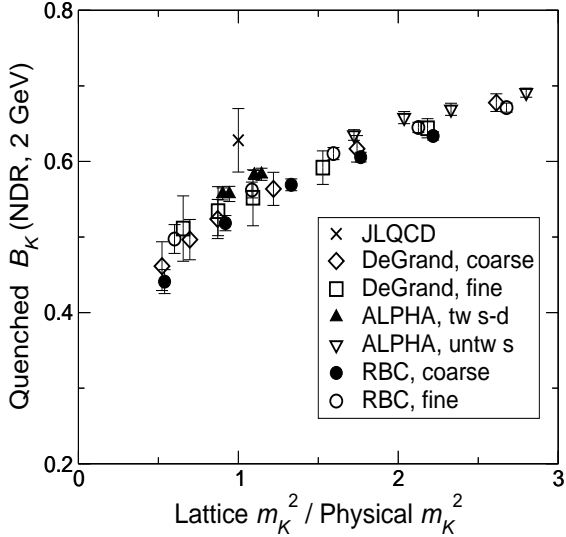}
\caption{\label{fig:QuenchBK} Quenched $\mathrm{B}_K$ as a function of 
$m_K^2$. Selected results, in legend order: 
\cite{Aoki:1997nr,DeGrand:2003in,Dimopoulos:2004xc,Noaki:Lat2004}.}
\end{figure}

In order for experimental measurements of $K^0-\overline{K^0}$ mixing
to constrain the CKM parameters $\overline{\rho}$
and $\overline{\eta}$, a
precise determination in QCD of the matrix element
\beq
\langle \overline{K^0} | O_{\Delta S = 2} | K^0\rangle 
\equiv \frac{8}{3} f_K^2 m_K^2 \mathrm{B}_K
\eeq
is necessary.  There have been many quenched calculations of $\mathrm{B}_K$;
a review was given recently in \cite{Gupta:2003hu}.  
Several new results have appeared since that review 
\cite{Becirevic:2004aj,Dimopoulos:2004xc,Noaki:Lat2004,Lee:2004qv,Gamiz:2004qx,Lellouch:Lat2004}.

\begin{table}[t]
\caption{Selected quenched $\mathrm{B}_K$ in the continuum limit
(an asterisk indicates a 2 point extrapolation).}
\label{tab:BK}
\begin{center}
\begin{tabular}{lcc} \hline
Ref.\ & fermions & $\mathrm{B}_K(\overline{\rm MS}, 2~\mathrm{GeV})$ \\ \hline
\cite{Aoki:1997nr} & staggered & 0.628(42) \\
\cite{AliKhan:2001wr}* & DWF & 0.575(20) \\
\cite{Dimopoulos:2004xc} & twisted $m$ &  0.592(16) \\
\cite{Noaki:Lat2004}* & DWF & 0.570(20) \\
\hline
\end{tabular}
\end{center}
\end{table}

The present situation of the quenched results,
is murky, but perhaps not dire.
Fig.~\ref{fig:QuenchBK} shows several different quenched results for
$\mathrm{B}_K$ plotted vs.\ lattice $K$ mass.  The cross is a staggered fermion
result in the continuum limit \cite{Aoki:1997nr}.  
The rest of the points are at a single lattice spacing, but are
instructive nevertheless.  One observes a downward curvature in $\mathrm{B}_K$
as one decreases the $K$ mass, as expected from chiral perturbation
theory.  Reports of linear-only mass dependence have been made based
on data which do not disagree with this plot: either the mass range
covered is too narrow or $m_K^2$ is not small enough to resolve the
curvature within statistical errors.

There is noticeable lattice spacing dependence in the RBC
domain wall fermion results indicating an upward trend in the continuum limit.
Indeed before comparing results too closely, discretization effects
must be removed.  Table~\ref{tab:BK} lists several results for $\mathrm{B}_K$
in the continuum limit (the DWF extrapolations both based on 2 lattice
spacings).  The disagreement between results is not too severe.
A clearer picture may emerge as studies with improved staggered 
fermions \cite{Lee:2004qv,Gamiz:2004qx} and overlap fermions
\cite{Lellouch:Lat2004} are pursued further.

\begin{figure}[t]
\vspace{5.0cm}
\includegraphics{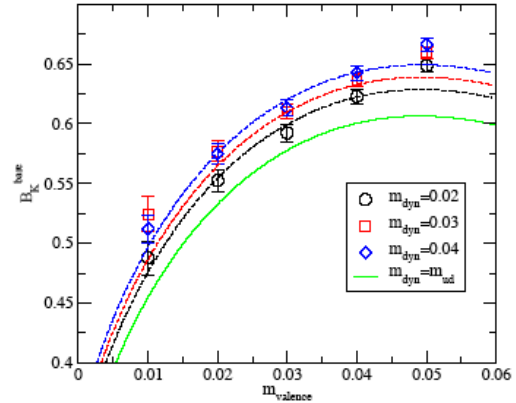}
\caption{\label{fig:RBCBK} Bare $\mathrm{B}_K$ vs.\ mass of valence
quark computed on the $n_f=2$ RBC
lattices with domain wall fermions \cite{Dawson:Lat2004}.
The physical result is obtained by interpolating to 
$am_{\rm val} = 0.022$, corresponding to $m_s/2$. The data at 
$am_{\rm val} = 0.01$ are discarded due to poor plateaux.  
Curves are from a fit to NLO PQ$\chi$PT.}
\end{figure}

More importantly for phenomenology, unquenched calculations for
$\mathrm{B}_K$ are underway.  Furthest along are the $n_f=2$ domain wall 
fermion results \cite{Dawson:Lat2004}.  Fig.~\ref{fig:RBCBK} 
shows $\mathrm{B}_K$
as a function of valence quark mass; different sea quark masses are
denoted by different symbols.  At $am_{\rm val}=0.022$, where physical
results are quoted, the data tend to decrease with the sea quark mass.
They quote $\mathrm{B}_K(\overline{\rm MS}, 2~\mathrm{GeV}) = 0.509(18)$
using only degenerate mass mesons.  If they use nondegenerate valence
quarks, and extrapolate the light mass to the physical limit they
find 0.496(17).  A correlated fit resolves the difference between these,
hinting that SU(3) breaking effects are becoming measurable.
These results are significantly lower than the quenched results in
Table~\ref{tab:BK}, and effect larger than anticipated from
Q$\chi$PT \cite{Sharpe:1996ih}.  The authors caution that simulations at
finer lattice spacing (as well as finite volume and lighter sea quark
mass) are necessary to estimate all the systematic effects.  

\begin{figure}[t]
\vspace{6.2cm}
\includegraphics{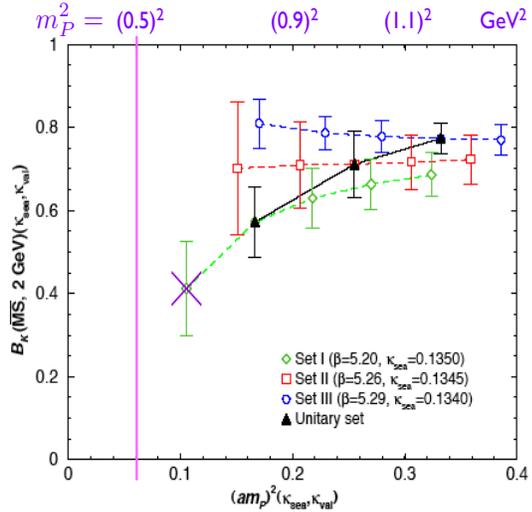}
\caption{\label{fig:UKQCDBK} $\mathrm{B}_K$ vs.\ mass of valence
pseudoscalar meson computed on the $n_f=2$ UKQCD
lattices with improved Wilson fermions \cite{Flynn:2004au}.
Increasing $\kappa$ implies decreasing $m_{\rm sea}$.
The lightest point is thrown out 
The vertical line denotes the physical $m_K^2$.}
\end{figure}

Unquenched data for $\mathrm{B}_K$ using improved Wilson fermions
recently appeared \cite{Flynn:2004au}.  The authors note that
the errors are too big and
the sea quark masses are too large for reliable extrapolation to
the physical masses.  However, they argue that their data
indicate $\mathrm{B}_K$ should decrease in the limit of physical 
sea quark masses
based on the observation of the partially quenched points
(shown in Fig.~\ref{fig:UKQCDBK}) decreasing at any given
value of $m_{\rm val}$ as the sea quark mass is decreased.
Given that the lightest $m_{\rm sea}\approx 1.4m_s$, I do not think 
the trend seen in Fig.~\ref{fig:UKQCDBK} necessarily continues all
the way to and through the chiral regime.

$\mathrm{B}_K$ is also being calculated on the $n_f=2+1$ MILC 
configurations; preliminary bare results were presented 
here \cite{Gamiz:2004qx}.  There is some concern that, after the
renormalization constant is computed, this result will agree with
the quenched JLQCD result \cite{Aoki:1997nr}
and consequently disagree sharply
with the $n_f=2$ RBC result \cite{Dawson:Lat2004}.  Such comparison,
however, is unjustified until the matching is done for \cite{Gamiz:2004qx}
and systematic uncertainties are better estimated.

\section{OUTLOOK}
\label{sec:concl}

\begin{table*}[t]
\caption{Summary of unquenched results for phenomenology. 
(See \cite{Aoki:2003xb,Yamada:2001xp} for $\mathrm{B}_B$ factors.)
On the left side, errors are quoted (stat)(sys),
on the right, (expt)(latt). 
An asterisk indicates a result is preliminary.
}
\label{tab:summary}
\begin{tabular}{clc|clcl}
\hline
quantity & Lattice 2004 result & Ref.\ &
quantity & Lattice 2004 result & Ref.\ & RPP 2004 \cite{Eidelman:2004wy} 
\\ \hline 
$f_\pi$ & 129.5(0.9)(3.5) MeV & \cite{Aubin:2004fs} &
    $|V_{us}|$ & ~0.2219(26) & \cite{Aubin:2004fs} & ~0.2200(26) \\[1mm]
$f_K$ & 156.6(1.0)(3.6) MeV & \cite{Aubin:2004fs} &
    $|V_{ub}|$ & ~$3.3(7)(4)\;10^{-3}$ & \cite{Aubin:2004fs} &
 ~$3.7(5) \;10^{-3}$  \\[1mm]
$f_D$ & 224${+10\choose -14}$(21) MeV & ~\cite{Simone:Lat2004}* &
    $|V_{cd}|$ & ~0.239(20)(26) &\cite{Aubin:2004ej} & ~0.224(12) \\[1mm]
$f_{D_s}$ & 263${+5\choose -9}$(24) MeV & ~\cite{Simone:Lat2004}* &
    $|V_{cs}|$ & ~0.97(2)(10) &\cite{Aubin:2004ej} & ~0.996(13) \\[1mm]
$f_{B_s}$ & 260(7)(28) MeV & \cite{Wingate:2003gm} &
    $|V_{cb}|$ & ~0.038(6)(1) & 
\cite{Shigemitsu:2004ft}*\cite{Okamoto:2004xg}* & ~0.0413(15) \\[1mm]
\hline
\end{tabular}
\end{table*}

Now that unquenched lattice calculations are
able to push to lighter quark masses, well inside the chiral
regime in the case of staggered fermions, predictions for
phenomenology are becoming more accurate.  Once the ambiguities
and uncontrolled errors of the quenching are behind
us, we can concentrate on more reliable estimates of the controllable
uncertainties.

Already unquenched lattice results for many important 
quantities are being obtained.  Fig.~\ref{fig:ratioplot} shows
the comparison for many quantities which are well known experimentally.
Table~\ref{tab:summary} summarizes the results presented here. 
The precision of the lattice calculations for $|V_{us}|$ and
$|V_{ub}|$ are comparable in precision to the Review of Particle 
Physics numbers \cite{Eidelman:2004wy}, and the heavy-light decay
constants are predictions.  I believe the systematic 
uncertainties in calculations of $f_B$ and $\mathrm{B}_K$ will soon 
reach the level where they too can be entered in the table.
These latter 2 are especially crucial to constraints on
the CKM parameters $\overline{\rho}$ and $\overline{\eta}$.

\section*{ACKNOWLEDGMENTS}

This talk profited from correspondence and conversation with
A.~Ali~Khan, I.~Allison, D.~Be\'cirevi\'c, C.~Bernard,
C.~Davies, C.~Dawson, T.~DeGrand, K.~Foley,
E.~Gamiz, S.~Gottlieb, A.~Gray, E.~Gulez,
S.~Hashimoto, Y.~Kayaba, A.~Kronfeld,
L.~Lellouch, C-J.D.~Lin, V.~Lubicz,
P.~Mackenzie, J.~Noaki, M.~Okamoto, F.~Palombi,
C.~Pena, O.~P\`ene,
S.~Sharpe, I.~Shipsey, J.~Shigemitsu, J.~Simone,
A.S.B.~Tariq, and N.~Yamada.
The author is supported by DOE grant DE-FG02-00ER41132.



\begin{thebibliography}{100}

\bi{Lubicz:Lat2004}
V.~Lubicz, 
arXiv:hep-lat/0410051.

\bi{Shipsey:Lat2004}
I.~Shipsey,  
arXiv:hep-lat/0411009.

\bibitem{Davies:2003ik}
C.~T.~H.~Davies {\it et al.}  
Phys.\ Rev.\ Lett.\  {\bf 92}, 022001 (2004).

\bibitem{Bernard:2001av}
C.~W.~Bernard {\it et al.}, 
Phys.\ Rev.\ D {\bf 64}, 054506 (2001).

\bibitem{Ishizuka:2002nm}
N.~Ishizuka, 
Nucl.\ Phys.\ Proc.\ Suppl.\  {\bf 119}, 84 (2003).

\bibitem{Sommer:2003wm}
R.~Sommer,
arXiv:hep-ph/0309320.



\bibitem{Wingate:2003gm}
M.~Wingate {\it et al.}, 
Phys.\ Rev.\ Lett.\  {\bf 92}, 162001 (2004).

\bibitem{Bernard:2002pc}
C.~Bernard {\it et al.},  
Phys.\ Rev.\ D {\bf 66}, 094501 (2002).

\bibitem{AliKhan:2000eg}
A.~Ali Khan {\it et al.},  
Phys.\ Rev.\ D {\bf 64}, 034505 (2001).

\bibitem{AliKhan:2001jg}
A.~Ali Khan {\it et al.},  
Phys.\ Rev.\ D {\bf 64}, 054504 (2001).

\bibitem{Aoki:2003xb}
S.~Aoki {\it et al.},  
Phys.\ Rev.\ Lett.\  {\bf 91}, 212001 (2003).

\bibitem{El-Khadra:1996mp}
A.~X.~El-Khadra, A.~S.~Kronfeld and P.~B.~Mackenzie,
Phys.\ Rev.\ D {\bf 55}, 3933 (1997).

\bi{Simone:Lat2004}
J.~Simone {\it et al.}, 
arXiv:hep-lat/0410030.

\bibitem{El-Khadra:1997hq}
A.~X.~El-Khadra {\it et al.}, 
Phys.\ Rev.\ D {\bf 58}, 014506 (1998).

\bibitem{Aubin:2004xd}
C.~Aubin and C.~Bernard, 
arXiv:hep-lat/0409027.

\bibitem{Wingate:2003ni}
M.~Wingate {\it et al.}, 
Nucl.\ Phys.\ Proc.\ Suppl.\  {\bf 129}, 325 (2004).

\bibitem{Gray:2004hd}
A.~Gray {\it et al.}, 
arXiv:hep-lat/0409040.

\bibitem{Yamada:2001xp}
N.~Yamada {\it et al.}, 
Nucl.\ Phys.\ Proc.\ Suppl.\  {\bf 106}, 397 (2002).

\bibitem{Kronfeld:2002ab}
A.~S.~Kronfeld and S.~M.~Ryan,
Phys.\ Lett.\ B {\bf 543}, 59 (2002).

\bibitem{Arndt:2004bg}
D.~Arndt and C.J.D.~Lin, 
Phys.\ Rev.\ D {\bf 70}, 014503 (2004).

\bibitem{Abada:2002xe}
A.~Abada {\it et al.}, 
Phys.\ Rev.\ D {\bf 66}, 074504 (2002).

\bibitem{Abada:2003un}
A.~Abada {\it et al.},
JHEP {\bf 0402}, 016 (2004).




\bibitem{Aubin:2004fs}
C.~Aubin {\it et al.} 
arXiv:hep-lat/0407028.

\bibitem{Aubin:2004he}
C.~Aubin {\it et al.}  
arXiv:hep-lat/0409041.

\bibitem{Marciano:2004uf}
W.~J.~Marciano,
arXiv:hep-ph/0402299.

\bibitem{Eidelman:2004wy}
S.~Eidelman {\it et al.}, 
Phys.Lett.\ B{\bf 592} (2004) 1.


\bibitem{Becirevic:2001xt}
D.~Be\'cirevi\'c {\it et al.}, 
JHEP {\bf 0204}, 025 (2002).


\bibitem{Shigemitsu:2004ft}
J.~Shigemitsu {\it et al.}, 
 arXiv:hep-lat/0408019.

\bibitem{Becirevic:1999kt}
D.~Be\'cirevi\'c and A.~B.~Kaidalov,
Phys.\ Lett.\ B {\bf 478}, 417 (2000).

\bibitem{Athar:2003yg}
S.~B.~Athar {\it et al.}, 
Phys.\ Rev.\ D {\bf 68}, 072003 (2003).

\bibitem{Foley:2002qv}
K.~M.~Foley and G.~P.~Lepage,
Nucl.\ Phys.\ Proc.\ Suppl.\  {\bf 119}, 635 (2003).

\bi{Foley:Lat2004}
K.~M.~Foley {\it et al.}, this conference.

\bibitem{Okamoto:2004xg}
M.~Okamoto {\it et al.}, 
arXiv:hep-lat/0409116.

\bibitem{Aubin:2004ej}
C.~Aubin {\it et al.}, 
 arXiv:hep-ph/0408306.

\bibitem{Becirevic:2004ya}
D.~Be\'cirevi\'c {\it et al.},
arXiv:hep-ph/0403217.

\bibitem{Hashimoto:1999yp}
S.~Hashimoto {\it et al.}, 
Phys.\ Rev.\ D {\bf 61}, 014502 (2000).



\bibitem{Gupta:2003hu}
R.~Gupta,
arXiv:hep-lat/0303010.

\bibitem{Becirevic:2004aj}
D.~Be\'cirevi\'c {\it et al.}, 
 arXiv:hep-lat/0407004.

\bibitem{Dimopoulos:2004xc}
P.~Dimopoulos {\it et al.}, 
 arXiv:hep-lat/0409026.

\bi{Noaki:Lat2004}
J.~Noaki, 
arXiv:hep-lat/0410026.

\bibitem{Lee:2004qv}
W.~Lee {\it et al.},
 arXiv:hep-lat/0409047.

\bibitem{Gamiz:2004qx}
E.~Gamiz {\it et al.},
 arXiv:hep-lat/0409049.

\bi{Lellouch:Lat2004}
L.~Lellouch, this conference.

\bibitem{Aoki:1997nr}
S.~Aoki {\it et al.}, 
Phys.\ Rev.\ Lett.\  {\bf 80}, 5271 (1998).

\bibitem{DeGrand:2003in}
T.~DeGrand, 
Phys.\ Rev.\ D {\bf 69}, 014504 (2004).

\bibitem{AliKhan:2001wr}
A.~Ali Khan {\it et al.}, 
Phys.\ Rev.\ D {\bf 64}, 114506 (2001).





\bi{Dawson:Lat2004}
C.~Dawson, 
arXiv:hep-lat/0410044.

\bibitem{Sharpe:1996ih}
S.~R.~Sharpe,
Nucl.\ Phys.\ Proc.\ Suppl.\  {\bf 53}, 181 (1997)
[arXiv:hep-lat/9609029].

\bibitem{Flynn:2004au}
J.~M.~Flynn, F.~Mescia, and A.S.B.~Tariq, 
arXiv:hep-lat/0406013.


\end{thebibliography}
\end{document}